\documentclass[%
 aip,
 amsmath,amssymb,
reprint,%
]{revtex4-2}
 
\usepackage{graphicx}
\usepackage{dcolumn}
\usepackage{bm}
\usepackage[utf8]{inputenc}
\usepackage[T1]{fontenc}
\usepackage{mathptmx}
\usepackage{etoolbox}
\usepackage{color}
\makeatletter
\def\@email#1#2{%
 \endgroup
 \patchcmd{\titleblock@produce}
  {\frontmatter@RRAPformat}
  {\frontmatter@RRAPformat{\produce@RRAP{*#1\href{mailto:#2}{#2}}}\frontmatter@RRAPformat}
  {}{}
}%

\makeatother
\begin{document}


\title{Dry launching of silica nanoparticles in vacuum}
\author{Ayub Khodaee}
 \affiliation{University of Vienna, Faculty of Physics, Vienna Center for Quantum Science and Technology (VCQ), Boltzmanngasse 5, A-1090 Vienna, Austria}
 \affiliation{Institute for Quantum Optics and Quantum Information, Austrian Academy of Sciences, A-1090 Vienna, Austria}

\author{Kahan Dare}%
 \affiliation{University of Vienna, Faculty of Physics, Vienna Center for Quantum Science and Technology (VCQ), Boltzmanngasse 5, A-1090 Vienna, Austria}
 \affiliation{Institute for Quantum Optics and Quantum Information, Austrian Academy of Sciences, A-1090 Vienna, Austria}

\author{Aisling Johnson}
 \affiliation{University of Vienna, Faculty of Physics, Vienna Center for Quantum Science and Technology (VCQ), Boltzmanngasse 5, A-1090 Vienna, Austria}
 
\author{Uro\v s Deli\'{c}}
 \affiliation{University of Vienna, Faculty of Physics, Vienna Center for Quantum Science and Technology (VCQ), Boltzmanngasse 5, A-1090 Vienna, Austria}

 \author{Markus Aspelmeyer}
 \email{markus.aspelmeyer@univie.ac.at}
 \affiliation{University of Vienna, Faculty of Physics, Vienna Center for Quantum Science and Technology (VCQ), Boltzmanngasse 5, A-1090 Vienna, Austria}
 \affiliation{Institute for Quantum Optics and Quantum Information, Austrian Academy of Sciences, A-1090 Vienna, Austria}
\date{\today}
\begin{abstract}

Clean loading of silica nanoparticles with a radius as small as $\sim 50$ nm is required for experiments in levitated optomechanics that operate in ultra-high vacuum. We present a cheap and simple experimental method for dry launching of silica nanoparticles by shaking from a polytetrafluoroethylene (PTFE) surface. We report on the successful launching of single silica nanoparticles with a minimum radius of $43$ nm, which is enabled by the low stiction to the launching surface. Nanoparticles with radii of $43$ nm and $71.5$ nm are launched with a high flux and small angular spread of $\sim \pm 10^\circ$, which allows for trapping in a tightly focused optical tweezer within a couple of minutes. The measured velocities are significantly smaller than $1~\text{m}/\text{s}$. The demonstrated launching method allows for controlled loading of dry nanoparticles with radii as small as $43$ nm into optical traps in (ultra-)high vacuum, although we anticipate that loading of smaller sizes is equally feasible.
\end{abstract}
\maketitle

\section{\label{sec:intro}Introduction}

Levitated particles and the manipulation of their center-of-mass motion in vacuum with optical tweezers, magnetic or Paul traps recently emerged as promising candidate systems to address questions in macroscopic quantum physics, thermodynamics in the quantum regime and for the search of new physics beyond the standard model \cite{gonzalez2021levitodynamics}. Optically levitated silica nanoparticles of around $100$ nm in diameter have recently been prepared in their motional quantum ground state~\cite{delic2020cooling, magrini2021real, tebbenjohanns2021quantum,ranfagniGS}. Furthermore, various proposals consider nanoparticles of similar sizes in order to create macroscopic superpositions or for matter-wave interferometry\cite{batemanmatterwave,neumeiersuperposition}. In addition, the absence of clamping allows for unique experimental possibilities such as dynamic and non-linear potential landscapes \cite{neumeiersuperposition,rondinKramers}, making levitated nanoparticles a valuable resource in technological applications such as force sensing~\cite{ranjit2015PRA,hebestreitsensing}. However, most experiments with levitated nanoparticles suffer from decoherence induced by gas collisions. This is in large part due to the loading mechanism, which is typically based on spraying nanoparticles from an aqueous solution with an ultrasonic nebulizer\cite{Nebulizer}. While this method is easy and cheap to implement, it contaminates the vacuum chamber and optical components within, and requires ambient pressure conditions to be carried out~\cite{gieseler2014dynamics}. Therefore, current efforts push towards technological improvements to routinely operate in ultra-high vacuum (UHV).

\begin{figure}
\includegraphics[width=\linewidth]{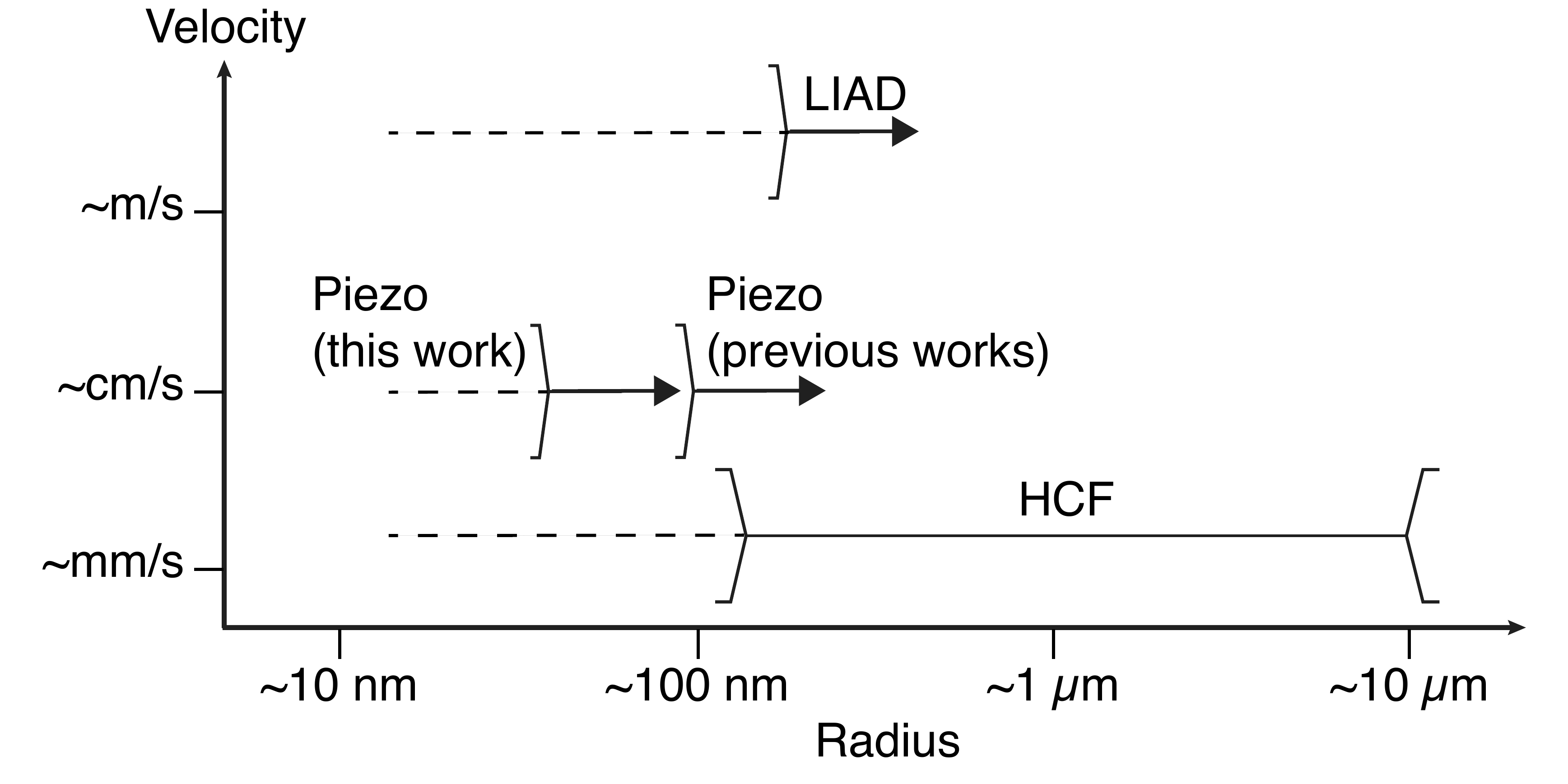}
\caption{\label{fig:SoA} State of the art launching mechanisms at low pressures (under ballistic conditions). Solid lines show the range of particle sizes that has been launched with a certain method, while the dashed lines present the potential for launching of smaller sizes. Laser-induced acoustic desorption (LIAD) is able to launch particles with small sizes, however with velocities higher than $1~\text{m}/\text{s}$ (in the ballistic regime)\cite{LIADAsenbaum,bykov2019direct,nikkhou2021direct}. Loading with hollowcore fibers (HCF) allows for the precise control of the nanoparticle velocities, and has been demonstrated for nanoparticles with a radius of $122.5~\text{nm}$\cite{rieser2020towards,lindner2021towards}. Shaking off nanoparticles from a substrate with the piezoelectric transducer (piezo) has been reported for particles with a radius of $88.5~\text{nm}$\cite{montoya2021scanning}. In this work we present the setup that is able to launch nanoparticles with a radius of $43~\text{nm}$. }
\end{figure}

An ideal loading strategy for the next generation of experiments in levitated optomechanics should be cost-efficient and adaptable to the variety of trapping techniques and experiments currently under development. Several loading mechanisms have been developed in this direction. Laser-induced acoustic desorption (LIAD) is able to load various particle sizes and materials into Paul traps at pressures down to $\sim 10^{-7}~\text{mbar}$~\cite{LIADAsenbaum,bykov2019direct, nikkhou2021direct}. Despite the large release velocity of particles, Paul traps create deep potential wells, thus allowing for trapping of fast particles in ultra-high vacuum by switching the trap on with precise timing \cite{bykov2019direct}. However, direct loading under similar conditions is difficult for the relatively shallow optical potentials without relying on the residual gas to slow down particles. Another strategy focuses on the transport of an externally trapped particle into the vacuum chamber, either by transporting the particle using a load-lock method~\cite{mestres2015cooling, calamai2021transfer} or with an optical conveyor belt through a hollowcore fiber~\cite{grass2016optical,rieser2020towards}. Although these methods were successful in handing over a nanoparticle into an optical tweezer, they come with additional optical components and lasers that add a significant overhead to experimental setups.


\begin{figure*}[ht]
\includegraphics[width=\linewidth]{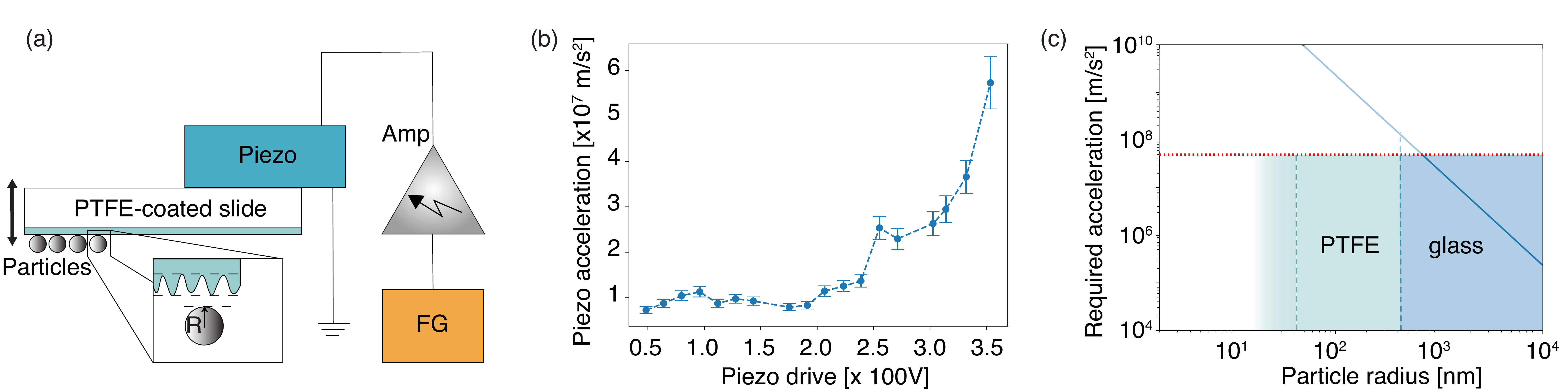}
\caption{\label{fig:setup} \textbf{(a)} The piezoelectric launching setup. A sinusoidal drive signal from the function generator (FG) is amplified with the high voltage amplifier by a factor of $100$ and applied to the piezo, which is clamped to the PTFE-coated slide. Nanoparticles with radius $R$ are launched from the surface of the oscillating PTFE coated slide. \textbf{(b)} The acceleration provided by the piezo at the piezo resonance frequency of $235.5$ kHz  as a function of the driving voltage. The acceleration increases drastically as the voltage increases. \textbf{(c)} The blue and the green shaded regions below the highest measured acceleration (red dotted line) represent the particle sizes that can be launched from the glass and the PTFE-coated slides, respectively. Acceleration required to launch nanoparticles from the glass slide as provided by the DMT model (blue solid line), crossing the maximum acceleration close to the launched nanoparticle size of $R=492.5$ nm (blue dashed line). Particles as small as $R=43$~nm were launched from the PTFE-coated slide (green dashed line).}
\label{fig:fig2setup}
\end{figure*}

In the pioneering work on optical tweezers, microparticles with a radius of $\sim 10~\mu\text{m}$ were shaken off glass substrates by using piezoelectric transducers (piezos)\cite{ashkin1971optical}. This method is cheap, simple, and versatile since it can in principle be used with any trap and in any environment. Although the particles are not transported deterministically into the final trap, the contamination of the environment is negligible in comparison to using a nebulizer. Recently, there has been a revived interest in launching small nanoparticles with this method \cite{park2017optical, montoya2021scanning, li2012fundamental,atherton2015sensitive, monge2017hybrid}. The main challenge is to overcome the strong stiction forces as the acceleration required to launch a nanoparticle scales with the inverse of the particle cross-section. In our experiment, we successfully release nanoparticles as small as $43$~nm in radius from a polytetrafluoroethylene (PTFE) coated substrate. In Figure~\ref{fig:SoA} we compare our result to the above-mentioned loading mechanisms in terms of velocity and particle sizes, highlighting the potential of the method developed in this work. In the following, we present the experimental setup and provide a detailed procedure to prepare particles on the substrate. We then characterize the flux, angular spread and velocity distribution of the launched nanoparticles. The analysis is based on images provided by scanning electron microscopy (SEM), which confirm the presence of single particles released from the launching substrate. 

\section{\label{sec:vdWmain}Van der Waals force}

The use of piezos to launch silica particles from glass surfaces has previously been realized in several experiments~\cite{ashkin1971optical, park2017optical, montoya2021scanning, li2012fundamental, atherton2015sensitive, monge2017hybrid}. However, high piezo drives are required to overcome the van der Waals stiction force between the particles and the surface, and the smallest launched particle had a radius of $88.5$ nm\cite{montoya2021scanning}. The Derjaguin-Muller-Toporov (DMT) theory predicts the stiction force between two particles with radii $R_1$ and $R_2$\cite{heim1999adhesion,derjaguin1975effect}:
\begin{equation}
    F_\mathrm{DMT} = 4\pi R_{\text{eff}} \gamma,
    \label{eq:DMT}
\end{equation}
where $\gamma$ is the effective solid surface energy between the two particles, and $R_{\text{eff}} = R_1R_2/(R_1+R_2)$. For a particle with the radius $R_1\equiv R$ deposited on a flat surface ($R_2 \to \infty$) it follows that the stiction force per mass (i.e. acceleration) increases as $F_\mathrm{DMT}/m\propto R^{-2}$. The effective surface energy of silica of $\gamma = 0.014~\text{J/m}^2$ has been measured between two silica microparticles\cite{heim1999adhesion}. The stiction force can effectively be reduced by using different materials with lower surface energies $\gamma$. In contrast to previous experiments, we use PTFE coated slides (coating thickness: $20~\mu\text{m}$) instead of glass slides, as the surface energy of PTFE is in general expected to be lower than for silica. However, there is little known about the stiction force between PTFE and silica under experimental conditions similar to the ones in our system. We note that higher surface roughness of the launching substrate might lead to significantly lower stiction force as it decreases the contact surface between the particles and the substrate \cite{liuroughness,barwell1975mechanics,johnson1998mechanics,fuller1975effect,persson2001effect}.

\section{\label{sec:setup}Launching setup}

The launching setup is shown in Figure~\ref{fig:setup}(a). The mechanical design, based on previous works \cite{park2017optical}, uses a clamped piezo to launch silica particles from a PTFE coated slide. One side of the PTFE coated slide is clamped to the piezo, while the particles are deposited on the other end by scraping them off a baked glass slide. Importantly, the process of baking a glass substrate with nanoparticles ensures that the stiction force dominates over any other attractive force, e.g. the capillary force\cite{paajanen2006experimental,heim1999adhesion}. The piezo is driven with a sinusoidal signal with frequency $\omega$, which is close to a piezo resonance such that the slide oscillates with maximum amplitude. Further information on the setup and the slide preparation can be found in Appendix \ref{sec:setupDetails}.  

The acceleration transferred to nanoparticles has to overcome the van der Waals force such that $a = F_{\text{DMT}}/m\propto R^{-2}$. This expression should be compared to the maximum acceleration provided by driving the piezo, $a_{piezo} =  \omega^2 \delta d$, where $\delta d$ is the substrate displacement. We choose the piezo resonance frequency at $235.5$ kHz, where the measured displacements (Appendix~\ref{sec:character}) correspond to accelerations of up to $\sim 6\times 10^7~\text{m}/\text{s}^2$ (Figure \ref{fig:setup}(b)). This sets the limit for the minimum particle radius that can be launched for a given substrate material and roughness (red dotted line, Figure~\ref{fig:fig2setup}(c)). Given the maximally achieved accelerations, following from the simple DMT model we should be able to launch particles from the glass slide down to $R\sim 630$~nm. In practice, we are able to launch a small number of particles with a radius of $492.5$ nm and no particles with a radius of $377.5$ nm, which approximately fits to the estimate. On the other hand, we are able to launch particles with a radius of $43$ nm from the PTFE surface. We conclude that the acceleration required to launch particles from the PTFE surface is more than two orders of magnitude smaller than for the glass slide. We note that nanoparticles with smaller radii were unavailable at the time of our measurements.

Using this method, we have successfully trapped $71.5$ nm particles in an optical trap (waist of $\sim 0.7 ~\mu$m) in the diffusive regime at pressures around $\sim 100$ mbar (Appendix~\ref{sec:trap}). For the remainder of the text we focus on the characterization of the launching method, namely on the flux, angular spread and launching velocities. We conduct all launching efforts at a pressure of $\sim 10^{-3}$ mbar, where the nanoparticles move in the ballistic regime (Appendix \ref{sec:ball}). Throughout our work we use silica nanoparticles with the nominal radii of $(43\pm 3)$ nm and $(71.5\pm 2)$ nm (Microparticles GmbH.). Note that these sizes are of particular interest as they correspond to particles that have recently been prepared in the motional quantum ground state \cite{delic2020cooling,magrini2021real,tebbenjohanns2021quantum}.

\section{Launching characterization}

\subsection{\label{sec:flux}Flux and angular spread}

It order to have a high repetition rate of experiments in levitated optomechanics, it is imperative to trap a nanoparticle in a reasonable time. Successful trapping of a nanoparticle depends strongly on the angular spread of launched particles and their flux, i.e. the number of particles passing through the trapping potential in a given time frame. In our consideration we focus on launching into an optical tweezer as it is presently the most common trapping method in experiments. The trap volume of an optical tweezer is given by $\sim \lambda^3$, where $\lambda$ is the laser wavelength. In our experiment we use $\lambda=1064$ nm, such that the flux numbers are reported as a number of particles passing through an area of $\sim 1~\mu \text{m}^2$ per second at a distance of $12$ mm from the launching substrate. 

\begin{figure}
\includegraphics[width=\linewidth]{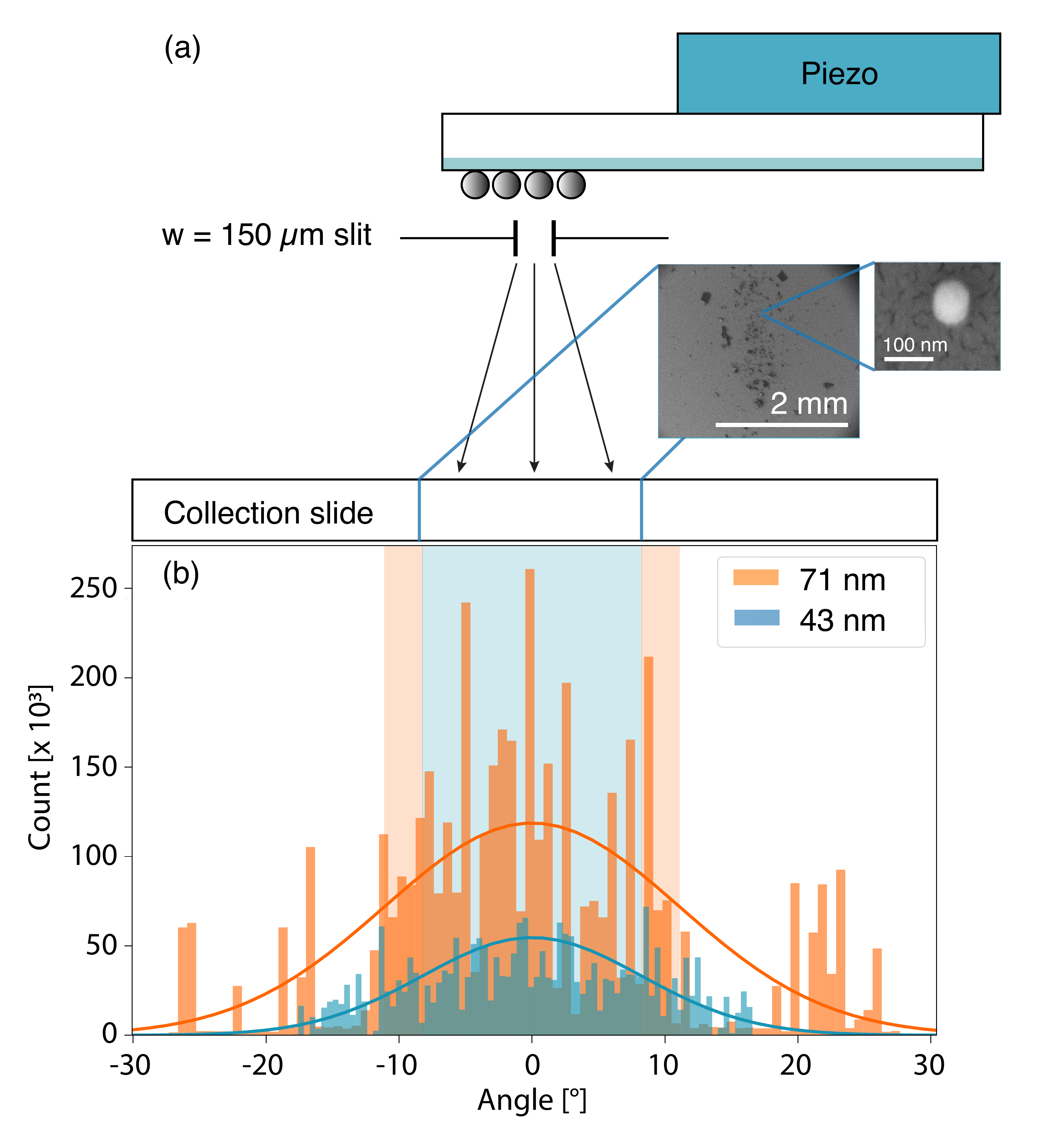}
\caption{\label{fig:flux} Measurement of the flux and angular spread of the launched nanoparticles. \textbf{(a)} The particles are launched along gravity onto the collection slide. A $150~\mu\text{m}$ wide slit selects a subset of the launched particles emulating a point source and is placed $12$ mm above the collection slide. Inset: SEM image showing single $43$~nm particles and clusters launched onto the collection slide. Zoom in: SEM image of a single $43$~nm nanoparticle on the collection slide. \textbf{(b)} The angular dependence of the launched particles is obtained by counting nanoparticles from the SEM images along the width of the slit. The total number of launched particles and the angular spread is lower for the $43$ nm particles, which is expected due to the higher van der Waals force to the substrate.}
\end{figure}

The setup to characterize the flux and the angular spread of the particles launched from the PTFE coated slide is shown in Figure~\ref{fig:flux}(a). A fraction of the launched particles and clusters travel through a long rectangular slit with a width of 150 $\mu$m and fall for approximately $12$ mm until they are deposited on a glass collection slide. We image the slide with a scanning electron microscope (SEM) with a magnification chosen such that a single particle covers $\sim 10$ pixels in each image. The total scanned area corresponds to the particles that have traversed the slit through an area of $(200\times150)~\mu\text{m}^2$ around its center. The images are then processed such that single particles and clusters are isolated from the background (Appendix~\ref{sec:sem}), however only the single particles are of interest to us for future experiments.

\begin{figure}
\includegraphics[width=\linewidth]{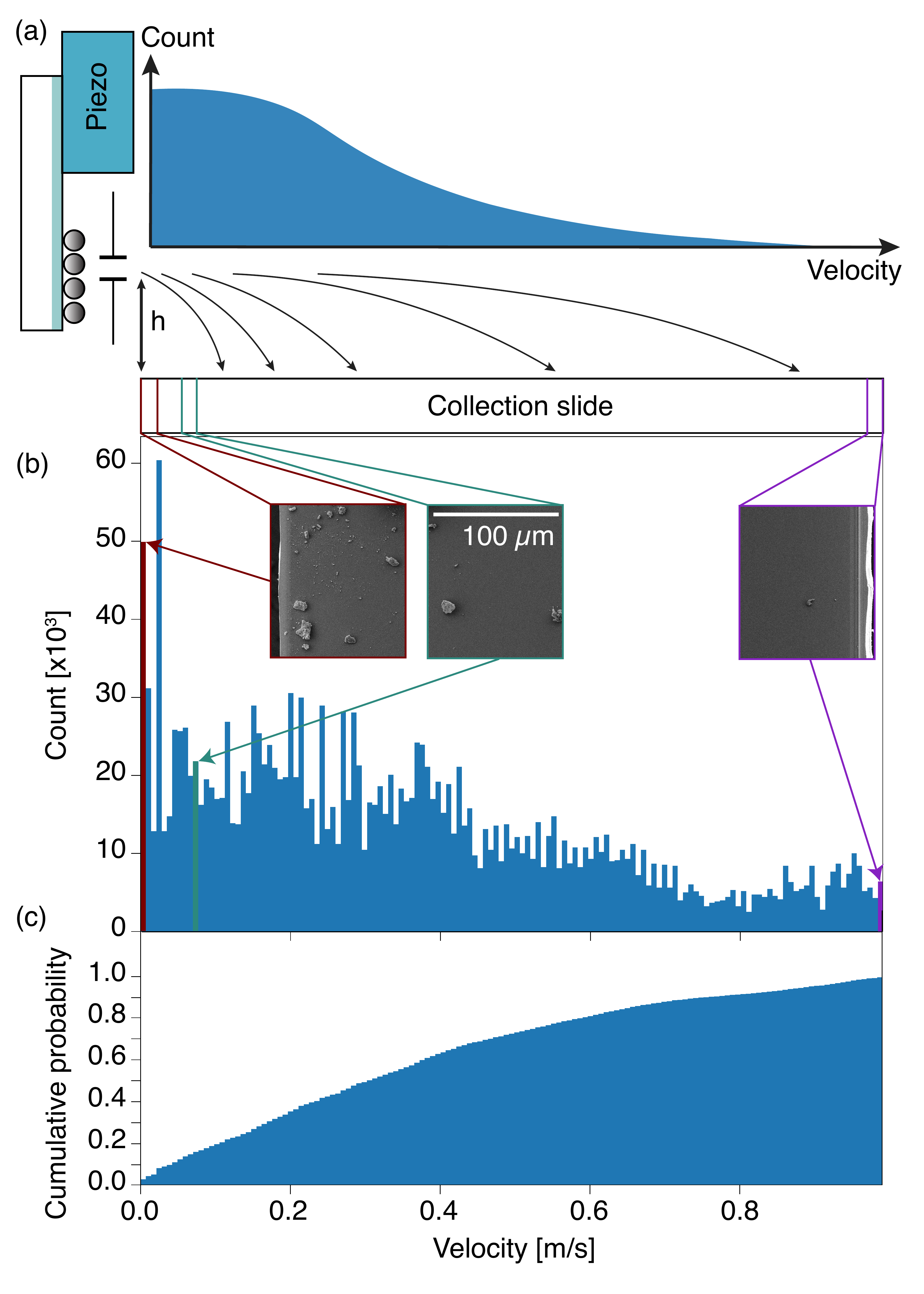}\caption{\label{fig:velocity} Measurement of the launching velocities. \textbf{(a)} Nanoparticles with a radius of $71.5$ nm are released along the horizontal axis and pass through a $150 ~\mu$m wide slit. The center of the slit is placed $h=11$ mm above a horizontally mounted collection slide, which is long enough to capture all particles that are launched with a velocity smaller than $1~\text{m}/\text{s}$. \textbf{(b)} We capture SEM images of the slide along its length and count the number of particles as a function of their distance from the slit. Insets: Three SEM photos are shown here from different sections of the slide as an example. The number of launched single nanoparticles and clusters clearly decreases the further away the image is taken is from the slit. \textbf{(c)} Cumulative probability of the launched nanoparticles as a function of the launched velocity.}
\end{figure}

The total single particle count is $4.56 \times 10^6$ for the $71.5$~nm particles and $2.3 \times 10^6$ for the $43$ nm particles in two hours of launching, giving a total flux for the given scanned area of $(0.021 \pm 0.005) ~\mu \text{m}^{-2}\text{s}^{-1}$ and $(0.011 \pm 0.005) ~\mu \text{m}^{-2}\text{s}^{-1}$, respectively. The higher count of the larger particles is expected as they require smaller accelerations to be launched. Although the flux is decreased for the smaller size, the high count suggests that launching of even smaller nanoparticles is feasible. However, this remains to be shown in future studies as particles with smaller radii were unavailable in the course of this work. We plot the particle count as a function of the angle in Figure~\ref{fig:flux}(b). We assume a Gaussian distribution of the collected nanoparticles and extract the angular spread of $(8.7\pm 0.5)^\circ$ ($(11.5\pm 0.5)^\circ$) for the $43$ nm ($71.5$ nm) nanoparticles, which is comparable to other techniques\cite{nikkhou2021direct}. The angular spread is larger for the larger particle size. We attribute this again to the smaller force required to launch the larger particles, which allows for a larger velocity component parallel to the launching slide.

\subsection{\label{sec:velocity}Launching velocities}

The trap depth of the optical tweezer is typically well below $10000~\text{K}$. In order to increase chances of trapping a nanoparticle flying through the trap volume, it is desirable to launch nanoparticles with energies smaller or on the order of the trap depth. We characterize the launching velocities with the setup presented in Figure~\ref{fig:velocity}. The launching assembly is now mounted vertically in order to release the nanoparticles in the horizontal direction. We mount a $150~\mu$m wide slit parallel to the launching substrate. The slit is placed at a height of $h = 11$ mm above a $5.0$ cm long glass slide, such that single $71.5$~nm particles and clusters that pass through the slit follow a parabolic trajectory and land at various distances from the base of the slit. We calculate the initial launching velocity for each nanoparticle from the measured travelled distance. 

The collection slide is imaged in steps along its length with the SEM. The image resolution is again chosen such that each particle is resolved by about 10 pixels. The insets in Figure \ref{fig:velocity} show the SEM images at three different distances along the collection slide, which clearly demonstrate that the particles and clusters are more densely concentrated in the vicinity of the launching assembly. The area scanned by the SEM in each image corresponds to particles passing through a $(150\times 120)~\mu\text{m}^2$ area of the slit. We count the particles as a function of the distance along the collection slide (Appendix~\ref{sec:sem}) and calculate the velocity distribution histogram. The total particle count is $9.46 \times 10^5$ in an hour of launching time. The total flux is then $(0.015\pm 0.005)~\mu\text{m}^{-2}\text{s}^{-1}$, which is consistent with the flux obtained in previous measurements. We observe that $30\%$ of the collected nanoparticles had a velocity smaller than $0.2~\text{m}/\text{s}$, allowing for direct cooling and trapping in an existing optical cavity \cite{vuletic2001three,delic2020cooling}. Around $17\%$ of nanoparticles have a velocity smaller than $0.07~\text{m}/\text{s}$ -- corresponding to the temperature of $T=mv^2/k_B\sim 1000$ K in a harmonic potential -- which could be captured in the optical tweezer with the help of methods developed for LIAD and hollowcore fibers\cite{bykov2019direct,lindner2021towards}. We note that the probability of launching slower nanoparticles might be increased by decreasing the piezo drive voltage, however this remains to be tested in future studies.


\section{Conclusion}

We present a cheap and simple method that launches dry nanoparticles off a PTFE coated substrate driven with a piezo in vacuum. We demonstrate successful launching of silica nanoparticles with a radius as small as $43$~nm, although from the measured particle flux we expect that launching of smaller particles is feasible. We show that particles with radii of $71.5$ nm and $43$ nm are launched with angular spreads of around $\pm 10^\circ$. Furthermore, $50\%$ of the launched particles have a launching velocity smaller than $\sim 0.3~\text{m}/\text{s}$. Based on our measurements a single particle would pass through the trap volume of an optical tweezer within $1-2$ minutes, depending on the size of the particle. The launching setup has a small footprint and can be easily implemented in any optical, magnetic or electric trap, thus enabling the next generation of experiments with levitated nanoparticles.

\begin{acknowledgments}
The authors are grateful for insightful discussions with M. Arndt, R. Quidant and M. Frimmer. We thank S. Puchegger, S. Loyer and B. Br\"auer for their help with the SEM and AFM measurements. This project was supported in part by the European Research Council (ERC 6 CoG QLev4G and ERC Synergy QXtreme), by the ERA-NET programme QuantERA under the Grants QuaSeRT and TheBlinQC (via the EC, the Austrian ministries BMDW and BMBWF and research promotion agency FFG), by the Austrian Science Fund (FWF) [I 5111-N],  and by the European Union’s Horizon 2020 research and innovation programme under Grant No. 863132 (iQLev).

\textit{Note added.} We recently became aware of similar work \cite{geracilaunching}.
\end{acknowledgments}

\section*{Data Availability Statement}

The data that support the findings of this study are available from the corresponding author upon reasonable request.

\section*{Author Declarations}
The authors have no conflicts to disclose.

\appendix

\section{\label{sec:setupDetails}Experimental setup and preparation of launching slides}

The launching assembly is shown in Figure \ref{fig:setup}(a). A cylindrical piezo (APC International, Ltd. part number 70-2221) is used throughout this project. The piezo was clamped on a PTFE coated glass slide (SPI supplies, part number 02285-AB). We cut the PTFE coated slides into pieces with dimensions $27~\text{mm} \times 50~\text{mm}$. The PTFE layer is $20~\mu$m thick. The function generator (FG) provides a continuous sine wave at the resonant frequency of the piezo, with maximum peak-to-peak voltages of $7$V. A high power voltage amplifier (Trek PZD350A) amplifies the signal by a factor of $100$. The amplifier can output signals with voltages up to $700$~V and frequencies up to $250$~kHz in bipolar mode, while the bandwidth is increased up to $350$~kHz in unipolar driving mode. As a result, the highest possible resonance frequency, i.e. $235.5$ kHz, was chosen for the experiments presented in this paper (see Appendix~\ref{sec:character} for more information on the resonances of the piezo).

The nanoparticles are produced by Microparticles GmbH and come in an aqueous solution with a nanoparticle weight concentration of $5\%$. We dilute the solution with isopropanol with a concentration of $5\%$ for $71.5$~nm and $2.5\%$ for $43$~nm particles. Around $50~\mu$l ($100~\mu$l) of the solution of the $71.5$~nm ($43$~nm) nanoparticles is deposited on a glass slide. We subsequently bake the slides at the temperature of $90^\circ$C for more than an hour in order to evaporate the water and isopropanol, as well as impurities off the porous silica particles. The dry particles are then scraped onto one end of the PTFE coated slide. This step is crucial as the low surface tension between the isopropanol-based solution and the PTFE coated slide prevents the direct deposition of the solution on the slide. The PTFE coated slide with the deposited particles is then baked in the oven at the same temperature for another $30$ minutes in order to get rid of the remaining impurities and possibly the absorbed water due to the humid environment during the scraping procedure. The humidity increases the stiction force \cite{paajanen2006experimental}, thus negatively affecting the flux of launched particles as the capillary force becomes non-negligible.

\section{\label{sec:character}Characterization of the setup}

\begin{figure}
\includegraphics[width=0.5\textwidth]{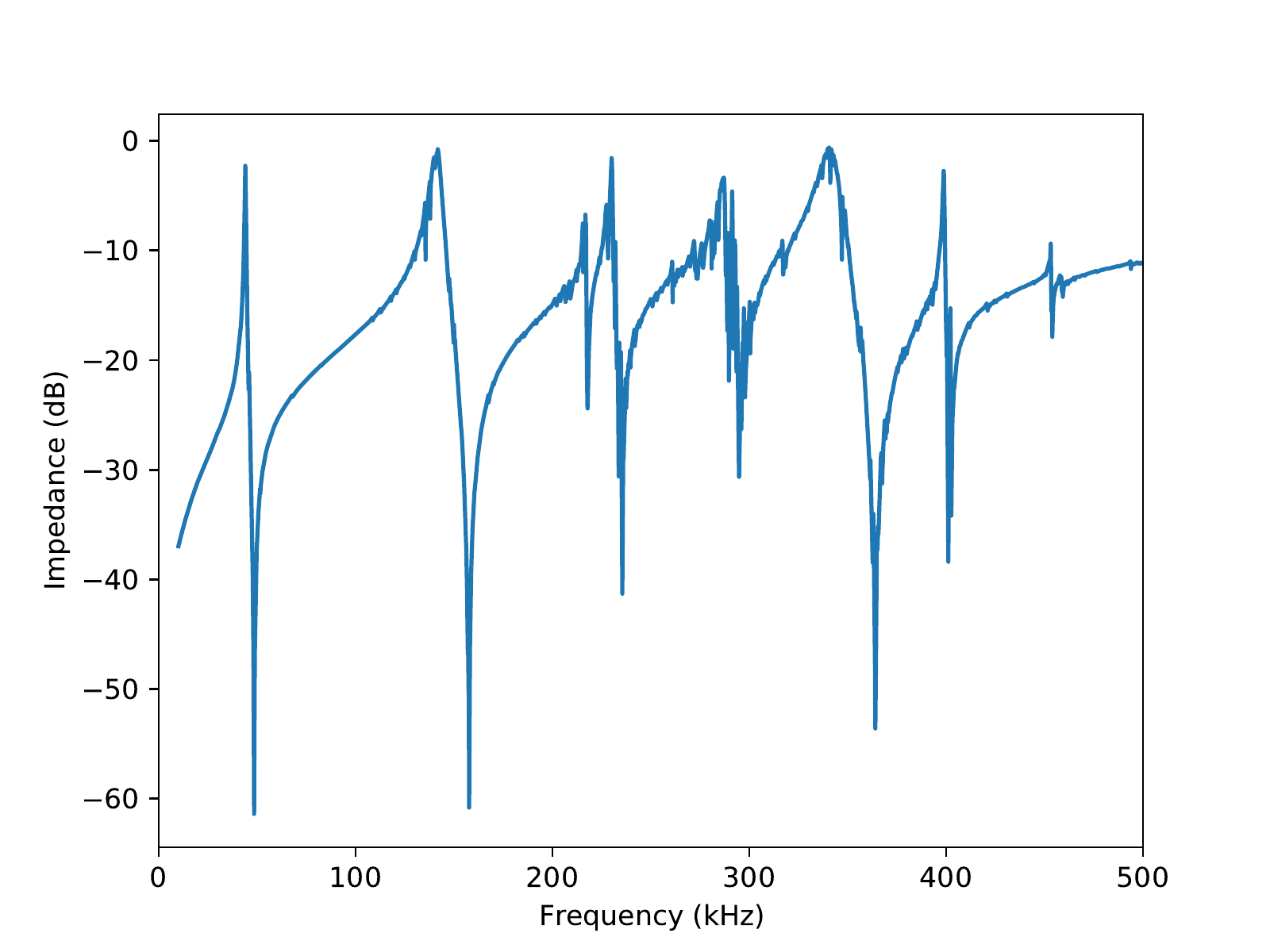}
\caption{\label{fig:S21} The S21 transfer function of the setup shown in the main text. The main frequencies are located at $155.8$, $235.5$, and $364.4$~kHz. Due to the amplifier's limited bandwidth, the resonance around 235.5~kHz was selected to drive the piezo.}
\end{figure}

The frequency response of the mounted piezo was measured using a network analyzer (impedance: $50~\Omega$) in order to find the optimal operating frequency. Figure \ref{fig:S21} shows the transmission S-coefficient $\text{S}_{21}$ as a function of the driving frequency, which is defined as $\text{S}_{21} = v_2/v_1$. The impedance $Z$ of the setup can be calculated as:
\begin{equation}
	Z = 50 \frac{1-S_{21, linear}}{ S_{21, linear}}\text{, where }S_{21, linear} = 10^{S_{21,dB}/20}.
\end{equation}
Here, $S_{21, linear}$ and $S_{21, dB}$ are the values of the $\text{S}_{21}$ transfer function in linear and logarithmic scales, respectively. The lower impedances correspond to the higher displacements of the piezo, thus resulting in higher accelerations to the particles. Its $\text{S}_{21}$ function is shown figure~\ref{fig:S21}. We detect several resonances at high frequencies: $155.8$ kHz, $235.5$ kHz, and $364.4$~kHz. As the acceleration depends on the frequency as $\propto \omega^2$, the largest frequency and displacement of the piezo would result in the highest acceleration. However, we select the resonance at the frequency of $235.5$~kHz as it provides large oscillation amplitudes and is within the bandwidth of the high-voltage amplifier.

\begin{figure}[h!]
\includegraphics[width=0.5\textwidth]{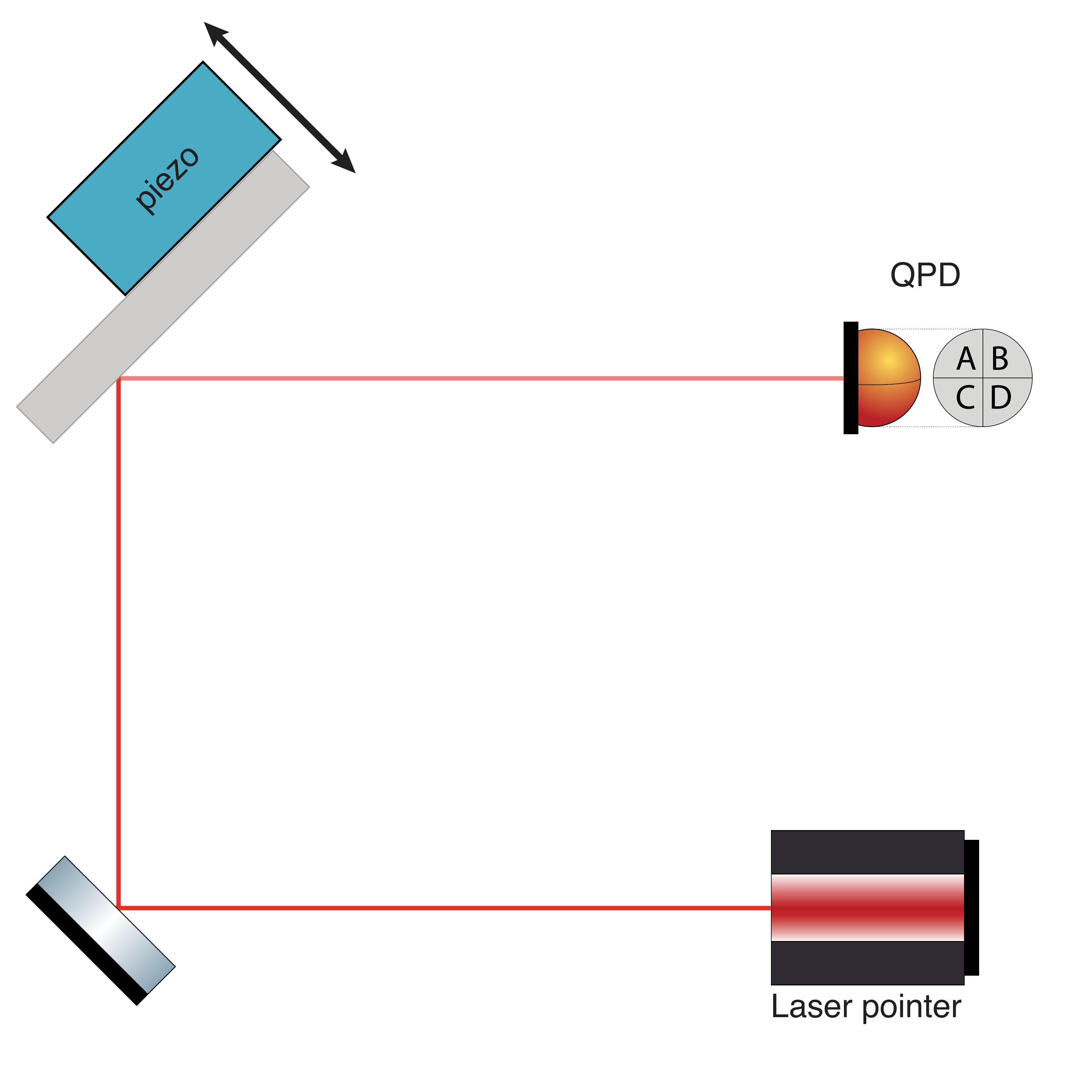}
\caption{\label{fig:character} The measurement of the displacement of the launching substrate. The laser light (power: $\sim 10$ mW) was reflected off the glass slide driven by the piezo and collected on a quadrant photodiode (QPD). The displacement of the launching substrate is detected by the displacement of the laser beam on the QPD.}
\end{figure}

We measure the displacement of the substrate by monitoring the shift of the laser beam that is reflected off the substrate (Figure~\ref{fig:character}). A laser pointer was reflected off a glass slide clamped to the piezo, and directed to a quadrant photodetector (QPD). The signal was calibrated by manually moving the substrate in steps of $0.5 ~\mu\text{m}$. The displacement was measured for the drive voltages of up to $350$~V at the drive frequency of $235.5$~kHz. We subsequently calculate the accelerations at the given drive frequency. We observe a nonlinear response of the piezo as we increase the voltage. Note that the calculation of the second derivate of the displacement over time provided a consistent measurement of the acceleration, which is the method of choice in Figure~\ref{fig:setup}(b) in the main text.

\section{\label{sec:ball}Validity of the ballistic approximation}

We consider a nanoparticle flying in a gas environment dominated by nitrogen. The nanoparticle of radius $R$ covering a distance $dx$ sweeps a cylinder of volume $\sigma dx$, where $\sigma=R^2\pi$ is the nanoparticle cross-section. The number of gas molecules $N$ it encounters is $n \sigma dx$, where $n=10^{19}~\text{molecules/m}^3$ is the molecular density of the gas. During the fall over a distance of $dx=11$~mm the nanoparticle will experience about $N=54$ collisions with gas molecules. In the worst case, a gas molecule imparts the momentum $2Nmv$ to the particle motion, where $v=\sqrt{2RT/M}$ is the most probable velocity of the gas molecules, $R=8.314 ~\text{J}\text{K}^{-1}~\text{mol}^{-1}$, $M=0.028~\text{kg}/\text{mol}$ and $m=4.65\times 10^{-26}$ kg is the molecular mass of nitrogen. All the collisions will maximally modify the nanoparticle momentum by $\Delta p\approx 10^{-21}~\text{kg}~\text{m}/\text{s}$, which in turn would slow down the nanoparticle by $\Delta v=\Delta p/2M_p\approx 7.5\times 10^{-4}~\text{m}/\text{s}$, where $M_p= 2.83\times 10^{-18}~\text{kg}$ is the mass of the nanoparticle. This change of velocity is an order of magnitude smaller than the width of individual bar ($7\times 10^{-3}~\text{m}/\text{s}$) in the velocity histogram. Given the negligible momentum kick transferred to the nanoparticles from the collisions with gas molecules, the assumption of the ballistic regime holds for this pressure and the particle sizes under consideration.

\section{\label{sec:sem}Post-processing of the SEM images}

The SEM images were taken along several adjacent parallel lines around the center of the collection slide to build a linear mosaic. The magnification of the SEM images was chosen such that each particle covers an area of around $3 \times 3 ~\text{pixels}^2$. The images were processed with a custom filter in order to increase the contrast of particles to the background, which is noisy mostly due to the uneven layers of gold coating applied to be used with the SEM. We first calculate the mean and the standard deviation of the values of the background pixels. We define all pixels as white if they have values higher than three to four standard deviations above the mean value of the background. Various thresholds result in different particle counts, which are then used to provide an error bar to the particle counting method. As a comparison, we analyzed the unfiltered images with the OpenCV python package, which detects circular objects and fits a circle to each particle. Both methods provide consistent numbers of detected nanoparticles. In the main text we choose to analyze the images based on the former method.

\section{\label{sec:trap}Optical trapping of nanoparticles}

\begin{figure}[h!]
\includegraphics[width=0.5\textwidth]{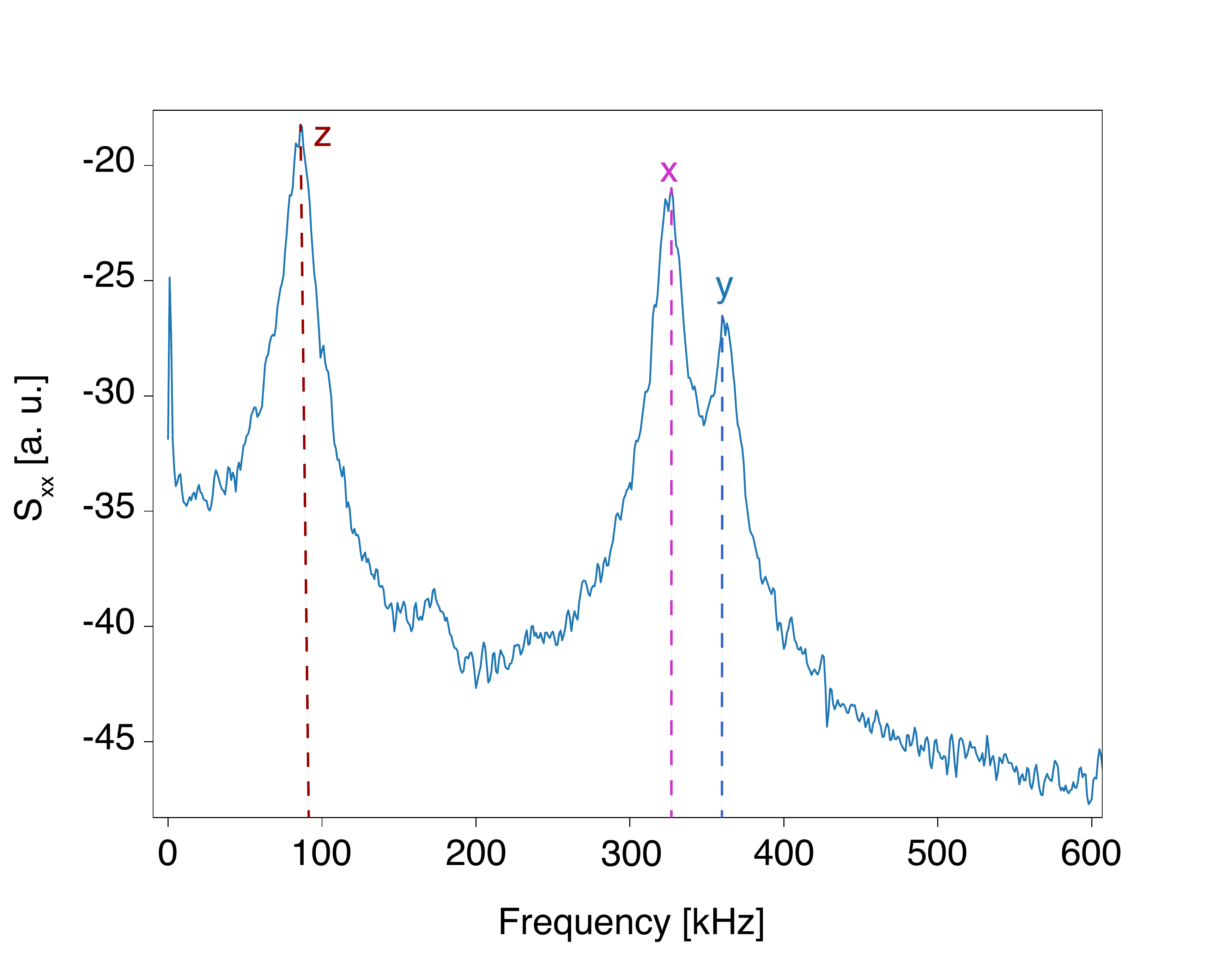}
\caption{\label{fig:PSD} The power spectral density of a single nanoparticle (radius: $71.5$ nm) that has been launched by shaking the piezo. The mechanical frequencies of the trapped particle along $x$, $y$ and $z$ directions are marked with pink, blue and red dashed lines, respectively. We note that the $z$ motion is along the optical axis of the tweezer. The power of the $1064$~nm laser before the chamber is $\sim$500 mW.}
\end{figure}

We have successfully trapped a single nanoparticle with a radius of $71.5$~nm  launched with our method. In the absence of optical cooling methods, we demonstrate this at the pressure of $\sim 100$~mbar, where the nanoparticle motion is diffusive and the gas friction helps slowing down the nanoparticle. The trap was generated by a $1064$nm laser beam focused to a waist of $\sim 0.7 ~\mu$m (numerical aperture of the trapping lens: $\text{NA}=0.77$). The piezo was driven for $30$ seconds with the voltage of $350$~V at the resonance frequency of $235.5$~kHz, after which the drive was turned off. A particle was trapped in the optical tweezer within minutes. The pressure in the vacuum chamber was then reduced to $1$~mbar in order to obtain good detection of the nanoparticle motion in the forward direction. The power spectral density of the particle motion is shown in Figure \ref{fig:PSD}. We note that we have trapped a single nanoparticle as trapping of clusters is unstable at lower pressures. We further verify this by loading a single nanoparticle with the nebulizer in the same experimental conditions, where the motional frequencies and the brightness on the camera are comparable to the case of loading by substrate shaking.

%

\end{document}